\documentclass[twocolumn,amsmath,floatfix]{revtex4}
\usepackage{latexsym}
\usepackage{amssymb}
\usepackage{pdfpages}
\usepackage{graphics}

\begin{document}
\newcommand{\ja}{Jakuba\ss a-Amundsen }
\newcommand{\bfx}{\mbox{\boldmath $x$}}
\newcommand{\bfq}{\mbox{\boldmath $q$}}
\newcommand{\bfnabla}{\mbox{\boldmath $\nabla$}}
\newcommand{\bfalpha}{\mbox{\boldmath $\alpha$}}
\newcommand{\bfsigma}{\mbox{\boldmath $\sigma$}}
\newcommand{\bfeps}{\mbox{\boldmath $\epsilon$}}
\newcommand{\bfA}{\mbox{\boldmath $A$}}
\newcommand{\bfP}{\mbox{\boldmath $P$}}
\newcommand{\bfe}{\mbox{\boldmath $e$}}
\newcommand{\bfn}{\mbox{\boldmath $n$}}
\newcommand{\bfW}{{\mbox{\boldmath $W$}_{\!\!rad}}}
\newcommand{\bfM}{\mbox{\boldmath $M$}}
\newcommand{\bfI}{\mbox{\boldmath $I$}}
\newcommand{\bfQ}{\mbox{\boldmath $Q$}}
\newcommand{\bfp}{\mbox{\boldmath $p$}}
\newcommand{\bfk}{\mbox{\boldmath $k$}}
\newcommand{\bfks}{\mbox{{\scriptsize \boldmath $k$}}}
\newcommand{\bfqs}{\mbox{{\scriptsize \boldmath $q$}}}
\newcommand{\bfxs}{\mbox{{\scriptsize \boldmath $x$}}}
\newcommand{\bfs}{\mbox{\boldmath $s$}_0}
\newcommand{\bfv}{\mbox{\boldmath $v$}}
\newcommand{\bfw}{\mbox{\boldmath $w$}}
\newcommand{\bfb}{\mbox{\boldmath $b$}}
\newcommand{\bfxi}{\mbox{\boldmath $\xi$}}
\newcommand{\bfzeta}{\mbox{\boldmath $\zeta$}}
\newcommand{\bfr}{\mbox{\boldmath $r$}}
\newcommand{\bfrs}{\mbox{{\scriptsize \boldmath $r$}}}

\renewcommand{\theequation}{\arabic{section}.\arabic{equation}}
\renewcommand{\thesection}{\arabic{section}}
\renewcommand{\thesubsection}{\arabic{section}.\arabic{subsection}}

\title{\Large\bf QED corrections to elastic electron-nucleus  scattering\\ beyond the first-order Born approximation}

\author{D.~H.~Jakubassa-Amundsen \\
Mathematics Institute, University of Munich, Theresienstrasse 39,\\ 80333 Munich, Germany}


\vspace{1cm}

\begin{abstract}  
A potential for the vertex and self-energy  correction is derived from the first-order Born theory.
The inclusion of this potential in the Dirac equation, together with the Uehling potential for vacuum polarization, allows for
a nonperturbative treatment of these QED effects within the phase-shift analysis.
Investigating the $^{12}$C and $^{208}$Pb targets, a considerable deviation of the respective cross-section change from the Born results is found, 
which becomes larger with increasing momentum transfer.
Estimates for the correction to the beam-normal spin asymmetry are also provided.
For the $^{12}$C nucleus, dispersion effects are considered as well.
\end{abstract}

\maketitle

\section{Introduction}

High-precision experiments with polarized or unpolarized electron beams \cite{Gr20,Au18} require an accurate knowledge of additional multiple photon processes which modify the Coulombic scattering
cross section.
To these belong, besides dipersion, the vacuum polarization and the vertex correction, renormalized by the self-energy and made infrared
finite by the soft bremsstrahlung.

For vacuum polarization it is well-known that the addition of the Uehling potential to the Coulombic target field $V_T$, which arises from the nuclear charge distribution, provides a nonperturbative consideration of this quantum electrodynamical (QED) effect \cite{Ue35,SM88}.
It is the first nonvanishing term in the decomposition of the vacuum loop in powers of $V_T$ \cite{Sh00}.
Indeed, if the Uehling potential were treated to first order \cite{Jaku21}, the respective cross-section modification would agree with Tsai's result \cite{T60,MT00} from the first-order Born approximation.  
However, the deviations from this Born prediction are, even for the $^{12}$C nucleus, formidable
in the vicinity of a diffractive cross-section minimum \cite{Jaku21b}.

The relation between the first-order Born amplitude and the underlying potential was recently applied in the context  of the contribution to the beam-normal spin asymmetry,
also known as Sherman function \cite{Mo64}, which results from dispersion.
In their method, Koshchii et al \cite{Ko21} constructed an absorptive potential from the respective Born amplitude,
to be included in the Dirac equation for the electronic scattering states, in order to provide a nonperturbative representation of the dispersive spin asymmetry.

In the present work this procedure is adopted for  generating a  potential $V_{\rm vs}$ for the vertex and self-energy (vs) correction from the respective first-order Born amplitude.
 Apart from the nonperturbative treatment of the  cross-section modifications induced by adding $V_{\rm vs}$ to $V_T$ in the Dirac equation, 
this allows for a
consistent estimate of the respective changes in the spin asymmetry.
By considering a light ($^{12}$C) and a heavy ($^{208}$Pb) target nucleus and electrons with energies between 1 MeV and 240 MeV,
the QED corrections and their dependence on the Coulomb distortion are investigated in a large region of momentum transfers.

The  paper is organized as follows. In section 2 the vs potential is derived. Results for the radiative modifications of the differential cross section and the spin asymmetry are provided in section 3 for the two target nuclei.
Concluding remarks follow (section 4).
Atomic units ($\hbar=m=e=1$) are used unless indicated otherwise.

\section{Theory}

In the Born approximation, the differential cross section for the elastic scattering of an electron into the solid angle $d\Omega_f$, which includes the radiative corrections to lowest order \cite{Lan}, is given by
$$\frac{d\sigma^{B1}}{d\Omega_f}=\frac{k_f}{k_i}\;\frac{1}{f_{\rm rec}}\;\sum_{\sigma_f}[ \;|A_{fi}^{B1}|^2$$
\begin{equation}\label{2.1}
 \;+\;2\mbox{ Re}\,\{A_{fi}^{\ast B_1}\,(A_{fi}^{\rm vac} + {A}_{fi}^{\rm vs}+A_{fi}^{\rm box})\} \;
 +\; \frac{d\sigma^{\rm soft}}{d\Omega_f}],
\end{equation}
where it is summed over the final spin polarization $\sigma_f$ of the electron. 
$A_{fi}^{B1}$ ist the first-order Born amplitude for potential scattering in the Coulombic target field $V_T$,
and $A_{fi}^{\rm vac}$ and $A_{fi}^{\rm box}$ are the lowest-order amplitudes for vacuum polarization \cite{MT00} and dispersion \cite{Sch55,Le56,FR74}, respectively.
 Recoil effects are
 considered  by the 
pre\-factor $f_{\rm rec}^{-1}$ \cite{Jaku21b}.
Here and in the following $k_i$ and $k_f$ denote the moduli of the initial and final electron momenta $\bfk_i$ and $\bfk_f$, respectively.

The lowest-order Born amplitude for the vertex correction, after eliminating the UV divergence by renormalizing with the help of the self energy, is given by \cite{BS19,Va00}
$$A_{fi}^{\rm vs}\;=\; F_1^{\rm vs}(-q^2)\;A_{fi}^{B1},$$
$$F_1^{\rm vs}(-q^2)\;=\;\frac{1}{2\pi c}\left[ \frac{v^2+1}{4v}\left(\ln \,\frac{v+1}{v-1}\right)\left( \ln\,\frac{v^2-1}{4v^2}\right)\right.$$
\begin{equation}\label{2.2}
+\;\frac{2v^2+1}{2v} \,\ln\,\frac{v+1}{v-1}
 -\,2\,+\;\frac{v^2+1}{2v}\left\{\mbox{Li }\left(\frac{v+1}{2v}\right)\right.
\end{equation}
$$ \left. \left.\;-\;\mbox{Li }\left(\frac{v-1}{2v}\right)\right\}\right]\;+\;\mbox{IR},$$
where 
 $q^2=(E_i-E_f)^2/c^2-\bfq^2$, with $\bfq=\bfk_i-\bfk_f$, is the squared 4-momentum transfer to the nucleus. 
$E_i$ and $E_f$ are the initial, respectively final, total energies  of the scattering electron.
Moreover, 
$v=\sqrt{1-4c^2/q^2}$ and Li$(x)=-\int_0^x dt \frac{\ln|1-t|}{t}\;$ is the Spence function \cite{T61,Va00}.
IR denotes the infrared divergent term. There is also a  magnetic contribution to $A_{fi}^{\rm vs}$ \cite{BS19}, which is tiny except for very low energies and which is omitted here.

The differential cross section for the soft bremsstrahlung reads in Born approximation 
\begin{equation}\label{2.3}
\frac{d\sigma^{\rm soft}}{d\Omega_f}\;=\;W_{fi}^{\rm soft}\;|A_{fi}^{B1}|^2
\end{equation}
with (correcting printing errors in \cite{MT00} and \cite{BS19})
$$ W_{fi}^{\rm soft}\,=\,-\frac{1}{\pi c}\left\{ 2 \,\ln\frac{2\omega_0}{c^2}\,+\,\frac{E_i}{k_ic}\,\ln\frac{c^2}{E_i+k_ic}\right. $$
$$\,+\,\frac{E_f}{k_fc}\,\ln \frac{c^2}{E_f+k_fc}
\,-\left[ 2\,\left(\ln \frac{2\omega_0}{c^2}\right)
\frac{v^2+1}{2v}\,\ln \frac{v+1}{v-1}\right.$$
$$+\,c^4\beta \,\frac{1-q^2/(2c^2)}{\zeta (\beta E_i-E_f)}\left( \frac{1}{4}\left( \ln \frac{E_i-k_ic}{E_i+k_ic}\right)^2 \!\!
-\,\frac{1}{4}\left( \ln \frac{E_f-k_fc}{E_f+k_fc}\right)^2 \right.$$
\begin{equation}\label{2.4}
\,+\,\mbox{Li}(1\,-\,\beta \,\frac{E_i-k_ic}{\zeta})
-\,\mbox{Li}(1\,-\,\frac{E_f-k_fc}{\zeta})
\end{equation}
$$ \left.\left. \left.\,+\,\mbox{Li}(1\,-\,\beta\,\frac{E_i+k_ic}{\zeta})\,-\,\mbox{Li}(1\,-\,\frac{E_f+k_fc}{\zeta})\right)\right] \right\}-\,2 \mbox{ IR},$$
introducing the cutoff frequency $\omega_0$ of the soft photons and the abbreviations
$$\beta\,=\,1\,-\,\frac{q^2}{2c^2}\,+\,\sqrt{-\frac{q^2}{c^2}\left( 1\,-\,\frac{q^2}{4c^2}\right)}$$
\begin{equation}\label{2.5}
\zeta\,=\,c^4\left[ \beta \left( 1\,-\,\frac{q^2}{2c^2}\right) -1\right]\frac{1}{\beta E_i-E_f}.
\end{equation}
The validity of (\ref{2.4}) for $W_{fi}^{\rm soft}$ is subject to the requirement that $\omega_0$ is not too small ($\frac{1}{\pi c} |\ln \frac{\omega_0}{c^2}|\ll 1$ \cite{Sh00}).
Due to mutual cancellations in (\ref{2.4}), a very large integration step number
for the Spence functions is necessary (some 50 000,
increasing with energy and angle).
For $-q^2/c^2\gtrsim 100$, the much simpler asymptotic formula for $W_{fi}^{\rm soft}$ can be used, as e.g. given in \cite{MT00} or \cite{Jaku21}.
Hard bremsstrahlung is disregarded in  (\ref{2.1}), since it is assumed that the resolution $\Delta E$ of the electron detector 
(which defines the upper limit of the photon frequency by $\omega_0=\Delta E$) is at most 1 MeV.

There is a simple connection between the first-order Born amplitude and the potential by which it is generated.
This is exemplified for the scattering amplitude $A_{fi}^{B1}$ which can be represented in terms of the nuclear
 charge form factor $F_L(|\bfq|)$ \cite{BD64},
\begin{equation}\label{2.6}
A_{fi}^{B1}(\bfq) \,=\, -\,\frac{2\sqrt{E_iE_f}}{c^2}\;\frac{Z}{\bfq^2}\;\left( u_{k_f}^{(\sigma_f)+}\;u_{k_i}^{(\sigma_i)}\right)\,F_L(|\bfq|),
\end{equation}
where $Z$ is the nuclear charge number and $u_{k_i}^{(\sigma_i)}, u_{k_f}^{(\sigma_f)}$ are, respectively, the free  4-spinors of the initial and final electronic states to the spin polarization $\sigma_i, \sigma_f$.
In turn, the form factor is related to the Fourier transform of the  target potential $V_T$,
\begin{equation}\label{2.7}
F_L(|\bfq|)\,=\,-\,\frac{\bfq^2}{4\pi Z}\int d\bfr\;e^{i\bfqs \bfrs} \,V_T(r).
\end{equation} 
This provides us with the basic relation between the
potential and the first-order Born amplitude,
$$V_T(r)\;=\;\frac{1}{(2\pi)^3}\int d\bfq\;e^{-i\bfqs\bfrs}\;A_{fi}^{B1}(\bfq)/A_0,$$
\begin{equation}\label{2.8}
A_0\;=\;\frac{\sqrt{E_iE_f}}{2\pi c^2}\;\left( u_{k_f}^{(\sigma_f)+}\;u_{k_i}^{(\sigma_i)}\right).
\end{equation}

For the construction of a nonperturbative theory, the IR contributions in (\ref{2.2}) and (\ref{2.4}) are omitted because it is known that they
cancel to all orders \cite{T61,Ye61}.
In order to derive the potential $V_{\rm vs}$ for the vertex and self-energy process, use is made of the
proportionality (\ref{2.2}) of its amplitude $A_{fi}^{\rm vs}$ to the scattering amplitude $A_{fi}^{B1}$.
Hence the application of (\ref{2.8}) yields
$$V_{\rm vs}(r)\;=\;\frac{1}{(2\pi)^3}\int d\bfq\;e^{-i\bfqs \bfrs}\;A_{fi}^{\rm vs}/A_0$$
\begin{equation}\label{2.9}
\approx\,-\,\frac{2Z}{\pi} \int_0^\infty d|\bfq|\;\frac{\sin(|\bfq|r)}{|\bfq|\,r}\;F_L(|\bfq|)\;F_1^{\rm vs}(-q^2).
\end{equation}
When performing the angular integration, the weak dependence of $F_1^{\rm vs}$ on $E_i-E_f$ (and hence on the scattering angle $\vartheta_f$) by means of recoil has been disregarded.

For a nonperturbative consideration of vacuum polarization and the vs correction, the Dirac equation
with the additional potentials is solved,
\begin{equation}\label{2.10}
\left[ -ic\bfalpha \bfnabla + \gamma_0 c^2 + V_T(r) +U_e(r)+V_{\rm vs}(r) \right]\psi(\bfr)=E\,\psi(\bfr),
\end{equation}
where $U_e$ is the Uehling potential \cite{Kl77} 
and $\bfalpha, \gamma_0$ refer to Dirac matrices.

\begin{figure}
\vspace{-1.5cm}
\includegraphics[width=11cm]{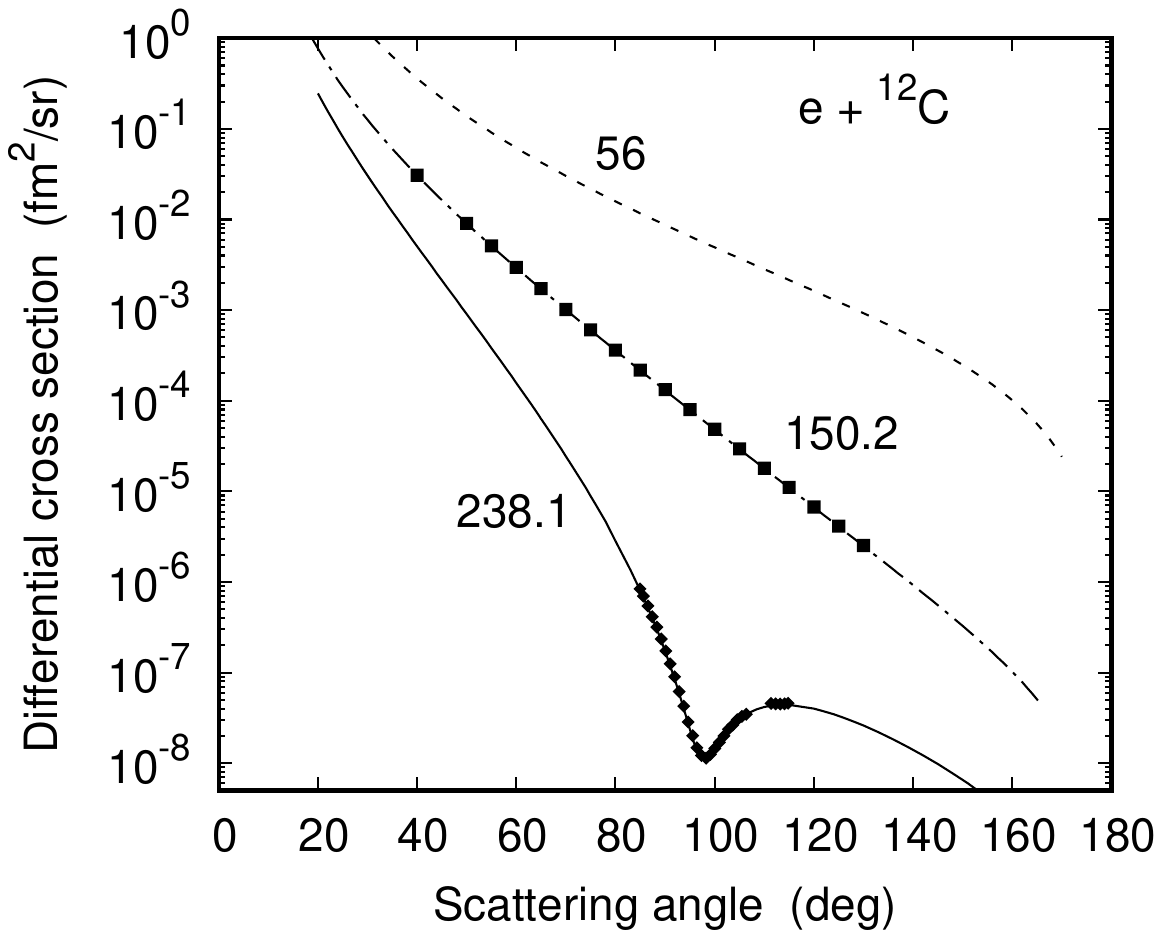}
\vspace{-0.5cm}
\caption
{
Differential cross section $\frac{d\sigma_{\rm coul}}{d\Omega_f}$ as a function of scattering angle $\vartheta_f$ for electrons of 56 MeV $(----$), 150.2 MeV $(-\cdot -\cdot -$) and 238.1 MeV (-------------) colliding with $^{12}$C. Also shown are the experimental data by Reuter et al ($\blacksquare$, \cite{Reu82}) at 150.2 MeV and by Offermann et al ($\blacklozenge$, \cite{Off91}) at 238.1 MeV.
}
\end{figure}

\section{Results}
\setcounter{equation}{0}

The Coulombic target potentials of $^{12}$C and $^{208}$Pb are generated from the Fourier-Bessel representation of the respective ground-state charge densities \cite{VJ}.
The electronic scattering state $\psi$ is expanded in terms of partial waves which, together with their phase shifts, are determined with the help of
 the Fortran code RADIAL of Salvat et al \cite{Sal}. 
Since the two additional potentials $U_e$ and $V_{\rm vs}$ are of long range
(as compared to the nuclear radius),  they require matching points between the inner and outer radial
solutions of the  Dirac equation of the order of 2000 fm.
The determination of the scattering amplitude involves weighted summations of the phase shifts \cite{Lan}, which are performed with the help of a threefold convergence acceleration \cite{YRW}.

In order to minimize the difference between the nonperturbative and the Born QED results, (\ref{2.1}) is  in the actual calculations modified  by including the Coulomb distortion throughout, as suggested by Maximon \cite{Ma69}.
This is done by replacing the Born amplitude $A_{fi}^{B1}$ by the exact Coulomb amplitude $f_{\rm coul}$, indicated in the replacement of $A_{fi}^{\rm vac},A_{fi}^{\rm vs}, d\sigma^{\rm soft}/d\Omega_f$ by $\tilde{A}_{fi}^{\rm vac}, \tilde{A}_{fi}^{\rm vs}, d\tilde{\sigma}^{\rm soft}/d\Omega_f$,
and is leading to
$$\frac{d\sigma^{B1-C}}{d\Omega_f} \,=\,\frac{k_f}{k_i}\,\frac{1}{f_{\rm rec}} \;\sum_{\sigma_f} [\;|f_{\rm coul}|^2 $$
\begin{equation}\label{3.1}
 +2\mbox{ Re}\, \{ f_{\rm coul}^\ast ( \tilde{A}_{fi}^{\rm vac}+\tilde{A}_{fi}^{\rm vs}+A_{fi}^{\rm box}) \}+\frac{d\tilde{\sigma}^{\rm soft}}{d\Omega_f}].  
\end{equation}
Hence, 
noting that $A_{fi}^{\rm vac}$ is, like $A_{fi}^{\rm vs}$, proportional to $A_{fi}^{B1}$ and
disregarding dispersion, the expression on the rhs of (\ref{3.1}) is proportional to the Coulombic cross section,
\begin{equation}\label{3.2}
\frac{d\sigma_{\rm coul}}{d\Omega_f}\,=\,\frac{k_f}{k_i}\,\frac{1}{f_{\rm rec}} \sum_{\sigma_f} |f_{\rm coul}|^2.
\end{equation}
The  scattering amplitude $f_{\rm coul}$ is obtained from the phase-shift analysis
relating to the potential $V_T$ \cite{Lan}.
Recoil is included in the phase-shift analysis in terms of a reduced collision energy $\sqrt{(E_i-c^2)(E_f-c^2)}$, in a similar way as done for excitation \cite{MG64}.

\begin{figure}
\vspace{-1.5cm}
\includegraphics[width=11cm]{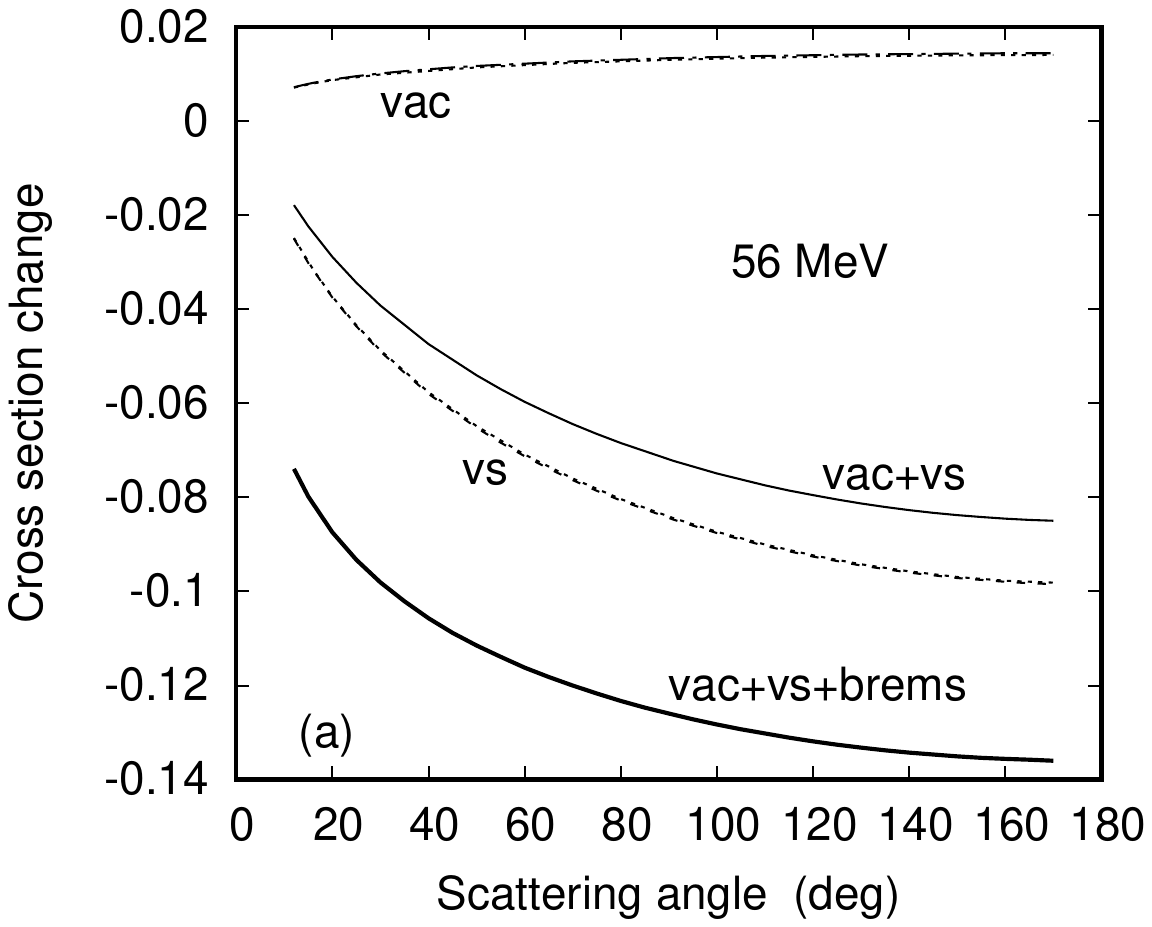}
\vspace{-1.5cm}
\vspace{-0.5cm}
\includegraphics[width=11cm]{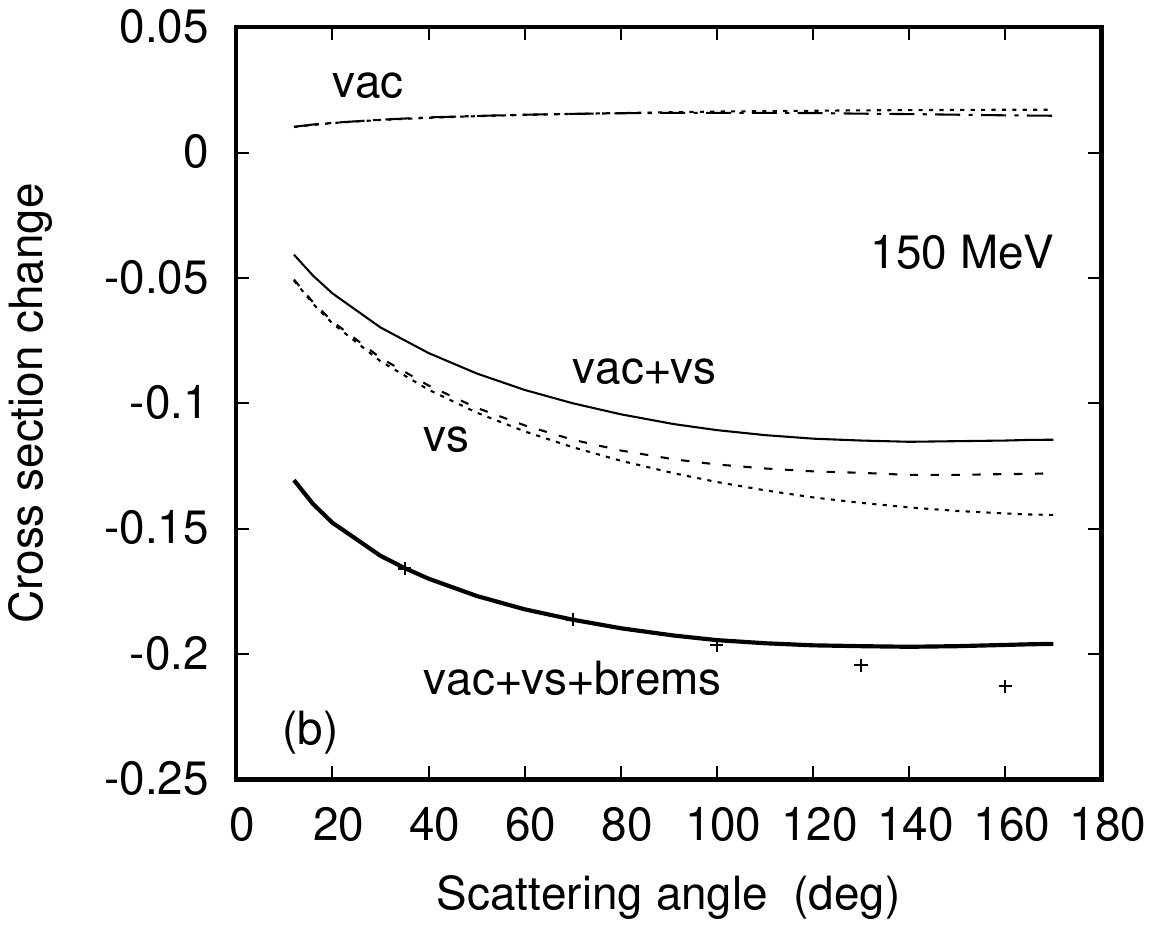}
\caption
{
Cross section change $\Delta \sigma^C$ in (a) 56 MeV and (b) 150 MeV $e+^{12}$C collisions as a function of scattering angle $\vartheta_f$.
Shown are the results for vacuum polarization $(-\cdot -\cdot -$), vertex and self-energy correction $(----$) and the consideration of both (----------, thin line), as well as the additional inclusion of the soft-bremsstrahlung contribution for $\omega_0=1$ MeV (------------, thick line).
Included are the Born results $\Delta \tilde{\sigma}^{\rm vac}$ for vacuum polarization and $\Delta \tilde{\sigma}^{\rm vs}$ for the vs correction $(\cdots\cdots$).
In Fig.2b, the crosses mark the sum of the combined QED corrections and dispersion (according to \cite{Jaku22}).
}
\end{figure}

On the other hand, the nonperturbative treatment of vacuum polarization and the vs process (leading to the scattering amplitude $f_{\rm vac+vs}$) results in the following expression for the differential cross section,
$$
\frac{d\sigma^{C}}{d\Omega_f} \,=\,\frac{k_f}{k_i} \,\frac{1}{f_{\rm rec}} \sum_{\sigma_f}\left[ \,|f_{\rm vac+vs}|^2\,+\,2\mbox{ Re } \{ f_{\rm coul}^\ast \,A_{fi}^{\rm box}\}\right. $$
\begin{equation}\label{3.3}
\left. +\; W_{fi}^{\rm soft}\;|f_{\rm vac+vs}|^2\right].
\end{equation}
In this prescription of the soft-photon cross section the fact has been accounted for that the cross section for emitting an additional soft photon during a certain scattering process
is given by the cross section for this scattering process times a factor which describes the attachment of one soft-photon line to the
respective diagram \cite{We65}.
This factorization holds as long as the  scattering process is undisturbed by this photon emission.
In particular, the photon energy has to be sufficiently low ($\omega_0 \ll E_i-c^2)$ and the change $\delta |\bfq|/|\bfq|$ of momentum transfer sufficiently small.
This restricts the scattering angle by means of \cite{Lan,Lo58} 
\begin{equation}\label{3.4}
\sin \frac{\vartheta_f}{2}\;\gg\;\frac{\omega_0c^4}{4\,E_i^3}. 
\end{equation}
For the present cases of interest, both conditions are well satisfied. In particular, one has for $\omega_0 \lesssim 1$ MeV and 
$E_i\gtrsim 50$ MeV the condition  $\vartheta_f \gtrsim 1^\circ$, or for $E_i-c^2 \gtrsim 1 $ MeV and an energy resolution of at most 1\% the requirement of $\vartheta_f \gtrsim 10^\circ$.
By using the Born factor $W_{fi}^{\rm soft}$ in (\ref{3.3}) the approximation is made that this soft-photon line corresponds to a free electron, in the same spirit as in the second-order Born representation of dispersion.

The effect of the QED and dispersion processes is illustrated by considering  the cross-section change, defined
with respect to the Coulombic cross section,
\begin{equation}\label{3.5}
\Delta \sigma \;=\;\frac{d\sigma/d\Omega_f}{d\sigma_{\rm coul}/d\Omega_f}\;-\,1,
\end{equation}
where in the two cross sections an additional averaging over the initial spin polarizaton $\sigma_i$ has to be made.
The cross section changes from the individual radiative processes are additive, such that
\begin{equation}\label{3.6}
\Delta \sigma^{B1-C} \,=\,\Delta \tilde{\sigma}^{\rm vac}
\,+\,\Delta \tilde{\sigma}^{\rm vs}\,+\,\Delta \sigma^{\rm box}\,+\,\Delta\tilde{\sigma}^{\rm soft},
\end{equation}
and
\begin{equation}\label{3.7}
\Delta \sigma^C\;=\;\Delta \sigma^{\rm vac+vs}\,+\,\Delta \sigma^{\rm box} \,+\,\Delta \sigma^{\rm soft},
\end{equation}
where the summands in (\ref{3.6}) and (\ref{3.7}) correspond to the contributions to $\Delta\sigma$
from the individual terms in (\ref{3.1}) and (\ref{3.3}), respectively.
It should be noted that $\Delta \sigma^{B1-C}$ --
without dispersion -- is approximately target-independent, since the Coulombic cross section drops out and recoil effects in vacuum polarization, $F_1^{\rm vs}$ and $W_{fi}^{\rm soft}$ are small.

\begin{figure}
\vspace{-1.5cm}
\includegraphics[width=11cm]{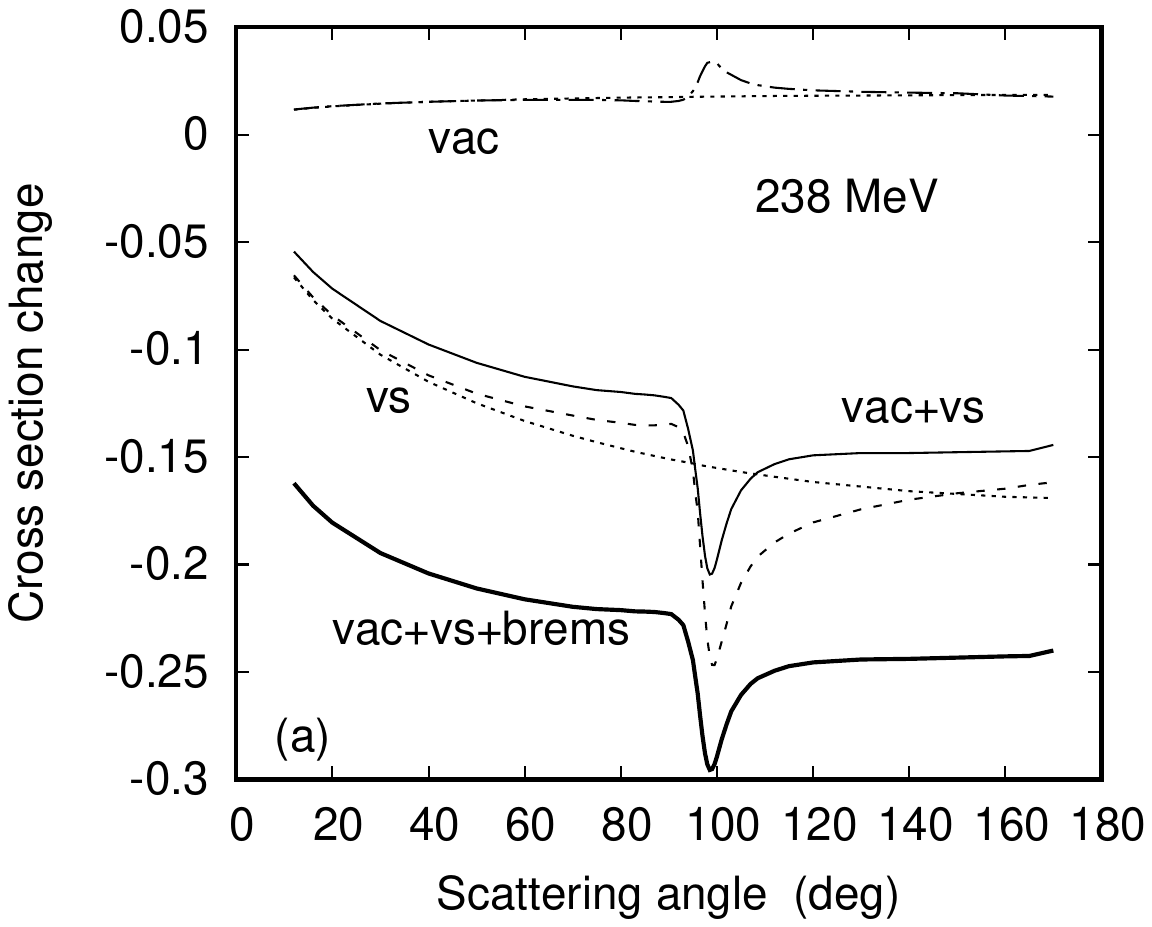}
\vspace{-1.5cm}
\vspace{-0.5cm}
\includegraphics[width=11cm]{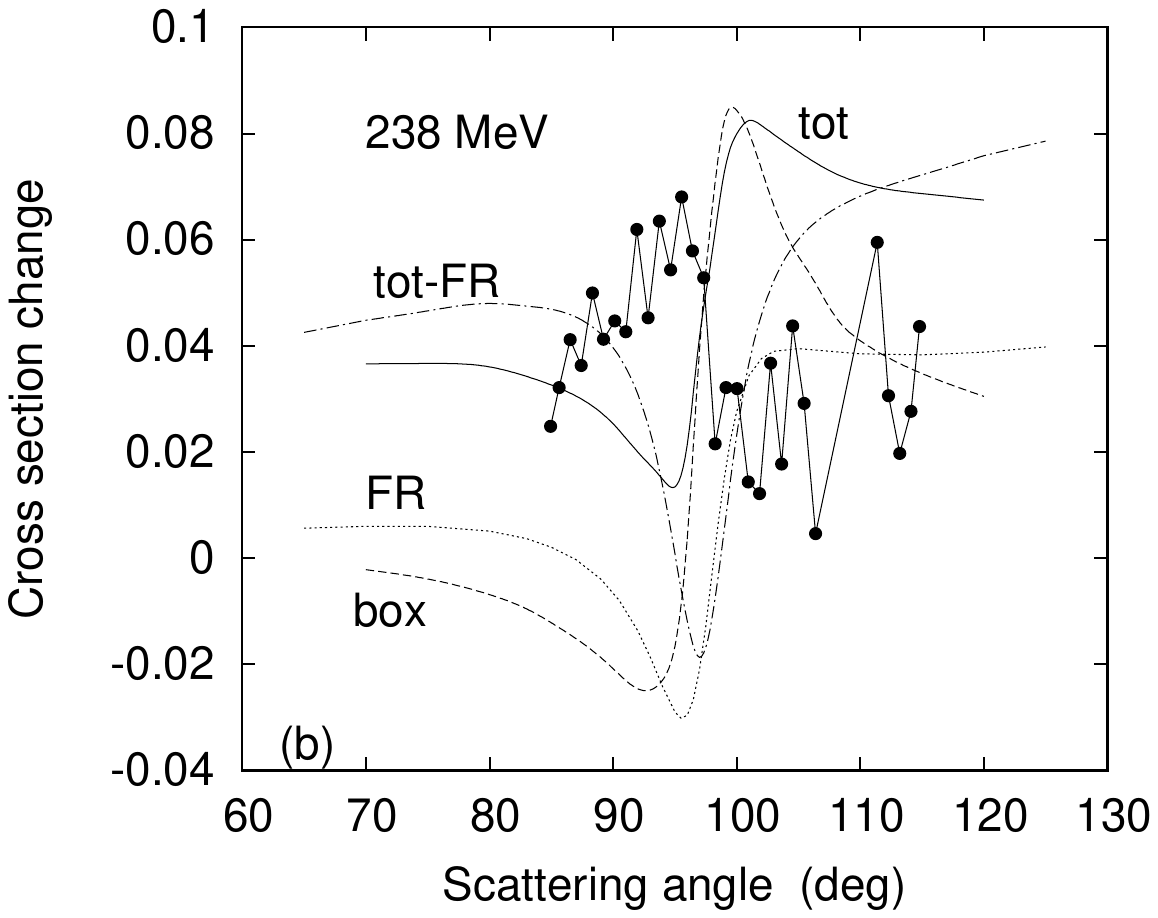}
\caption
{
Cross section change $\Delta \sigma^C$  in 238.1 MeV $e+^{12}$C collisions as a function of scattering angle $\vartheta_f$.
In (a), the lines have the same meaning as in Fig.2. In (b), $\Delta \sigma^{\rm box}$ calculated with the three transiently excited states ($----$) and with the Friar-Rosen theory $(\cdots\cdots$) are shown,
 as well as $\Delta\sigma^{\rm box} + [\Delta\sigma^C - \Delta \sigma^{B1-C} ]$ (-----------, present results according to \cite{Jaku22}, $-\cdot -\cdot -$ with the Friar-Rosen theory for $\Delta\sigma^{\rm box}$).
Included is the relative deviation of the experimental cross section from the Coulombic result ($\bullet$, connected by lines \cite{Off91}).
For $\omega_0$, the experimental value (0.05 MeV) is used, corresponding to the energy resolution of 0.02\%.
}
\end{figure}

\subsection{The $^{12}$C nucleus}

The angular distribution of the Coulombic cross section is shown in Fig.1 for the collision energies 56, 150.2 and 238.1 MeV.
Whereas at the two lower energies there is a monotonous decrease with scattering angle,
a diffraction minimum exists near $100^\circ$ for the 238.1 MeV electron impact. There is good agreement with the available experimental data \cite{Reu82,Off91}, which are corrected for global QED effects.

Fig.2 displays the corresponding cross section changes (\ref{3.7}) by the QED effects and dispersion in comparison with the (Coulomb-distorted)
Born approximation (\ref{3.6}). Apart  from showing the combined influence of these effects,
the vacuum polarization as well as the vs effect and their supersposition are provided separately. This is done by only retaining the respective potentials in (\ref{2.10}) or the corresponding perturbative parts in (\ref{3.1}).
For 56 MeV impact (Fig.2a), the deviations from the Born results are, as expected for this light nucleus, extremely small.
Bremsstrahlung enhances the cross-section change, dependent on the cut-off frequency $\omega_0$.
If not stated otherwise, $\omega_0=1 $ MeV is taken throughout,
which is well below the first excited state at 4.4 MeV for $^{12}$C or at 2.6 MeV for $^{208}$Pb.
The contribution $\Delta \sigma^{\rm box}$ is tiny at this energy (decreasing from $\sim -10^{-4}$ to $\sim - 10^{-3}$ with angle), and its inclusion is not visible in the graph.

At 150 MeV (Fig.2b) the influence of the QED effects on the cross section is notably smaller at the larger angles when the vs potential is considered nonperturbatively, than when predicted by the Born approximation.
Also a dispersion effect is peceptible at angles beyond $70^\circ$.

The situation is different at 238.1 MeV (Fig.3).
While the Born results for vacuum polarization or the vs  effect still have a monotonous angular dependence,
 the nonperturbative results mimic the presence of the 
diffraction minimum in $d\sigma_{\rm coul}/d\Omega_f$ by a resonance-like structure.
The comparision with Fig.2 indicates that the global strength of the vs correction  increases with $E_i$, whereas vacuum polarization remains at $1-2\%$.

\begin{figure}
\vspace{-1.5cm}
\includegraphics[width=11cm]{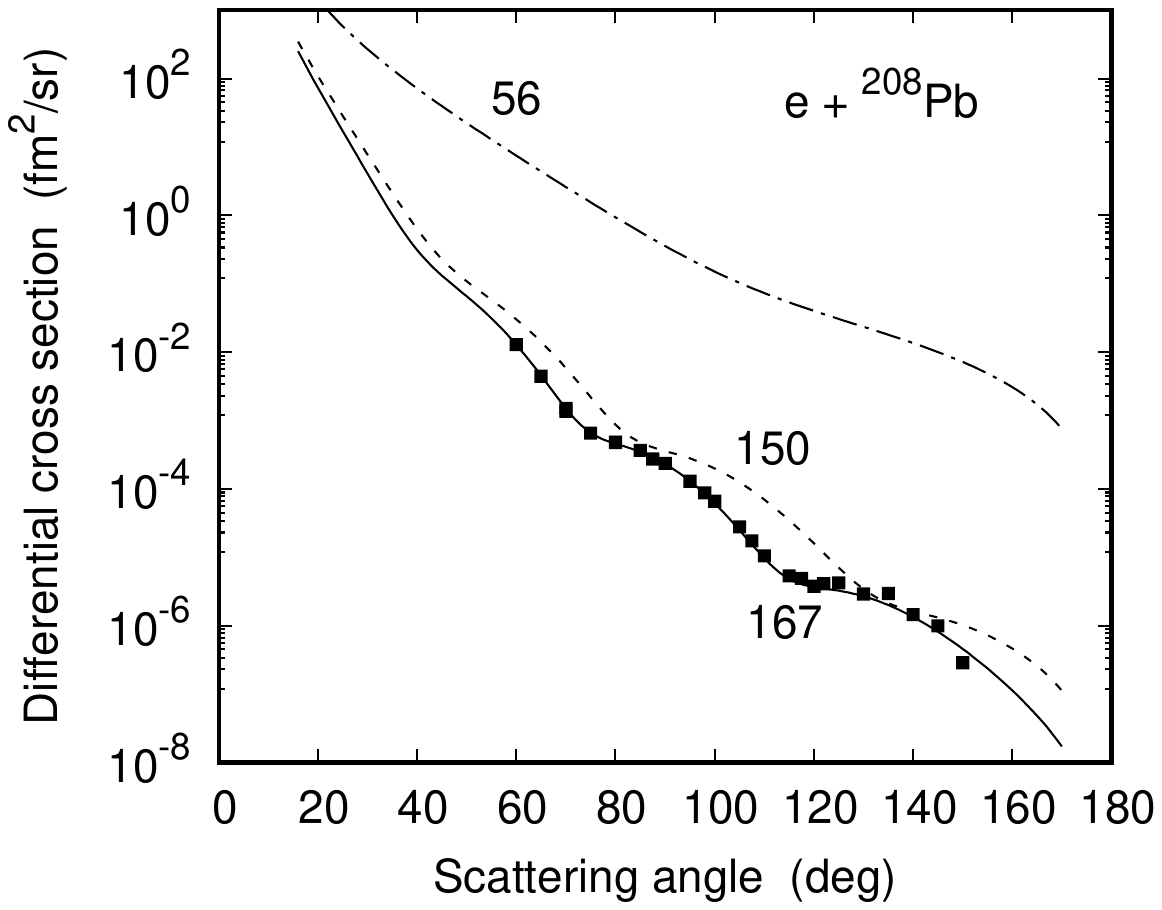}
\vspace{-0.5cm}
\caption
{
Differential cross section $\frac{d\sigma_{\rm coul}}{d\Omega_f}$ for electrons of 56 MeV $(-\cdot -\cdot -$), 150 MeV $(----)$ and 167 MeV (---------)\\ colliding with $^{208}$Pb as a function of scattering angle $\vartheta_f$. Included are the relative experimental data by Friedrich and Lenz ($\bullet$, \cite{Fr72}) at 167 MeV.
}
\end{figure}

The dispersion correction
 $\Delta \sigma_{\rm box}$ is displayed in Fig.3b. It is estimated in the second-order Born approximation by considering
three dominant transiently excited nuclear states 
of low angular momentum \cite{Jaku22}. It is seen that the resonant-like structure is also present in this  correction.
For comparison,  the result from the Friar-Rosen theory \cite{FR74} is included, which employs a closure approximation by setting all nuclear excitation
energies to the fixed value of 15 MeV \cite{Jaku21}.
We recall that in the experimental data by Offermann et al \cite{Off91} only a smooth background from the QED effects had been subtracted.
In order to account for the influence of the structure in the nonperturbative approach, it is isolated by forming the difference
between $\Delta\sigma^C$ and $\Delta \sigma^{B1-C}$.
This difference is  added to the dispersion correction, which is also not considered in the data, and the result is included in Fig.3b for
the two theories of $\Delta \sigma^{\rm box}$. Comparison is made with the experimental cross section change, obtained from (\ref{3.5}) by identifying $d\sigma/d\Omega_f$ with the cross section measurements.
One has to keep in mind that these so generated data points depend crucially on the way
how recoil is incorporated into the Coulombic result.
For a light target like $^{12}$C at such a high impact energy even the recoil prefactor $k_f/(k_i f_{\rm rec})$ in (\ref{3.2}) reduces the cross section by 5\%,
apart from the shift in angle by the reduced collision velocity \cite{Jaku21b}.
Although, like for $\Delta \sigma^{\rm box}$ alone, the deviation between theory and experiment around $95^\circ$ persists,
the consideration of the oscillatory behaviour of the QED corrections improves the agreement below $100^\circ$.
The Friar-Rosen theory is inferior in reproducing the experimental data, as is also known from its performance at higher energies \cite{Jaku22}.

\begin{figure}
\vspace{-1.5cm}
\includegraphics[width=11cm]{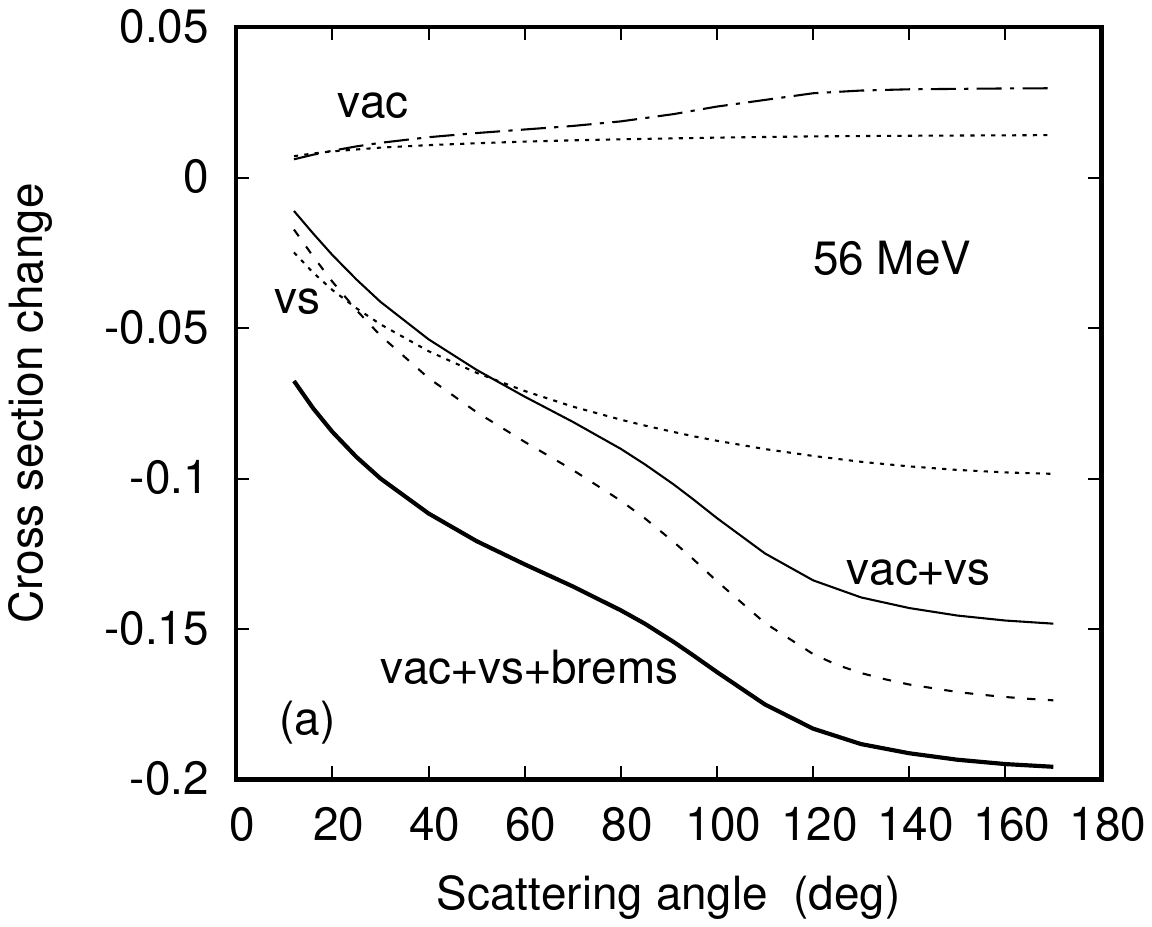}
\vspace{-1.5cm}
\vspace{-0.5cm}
\includegraphics[width=11cm]{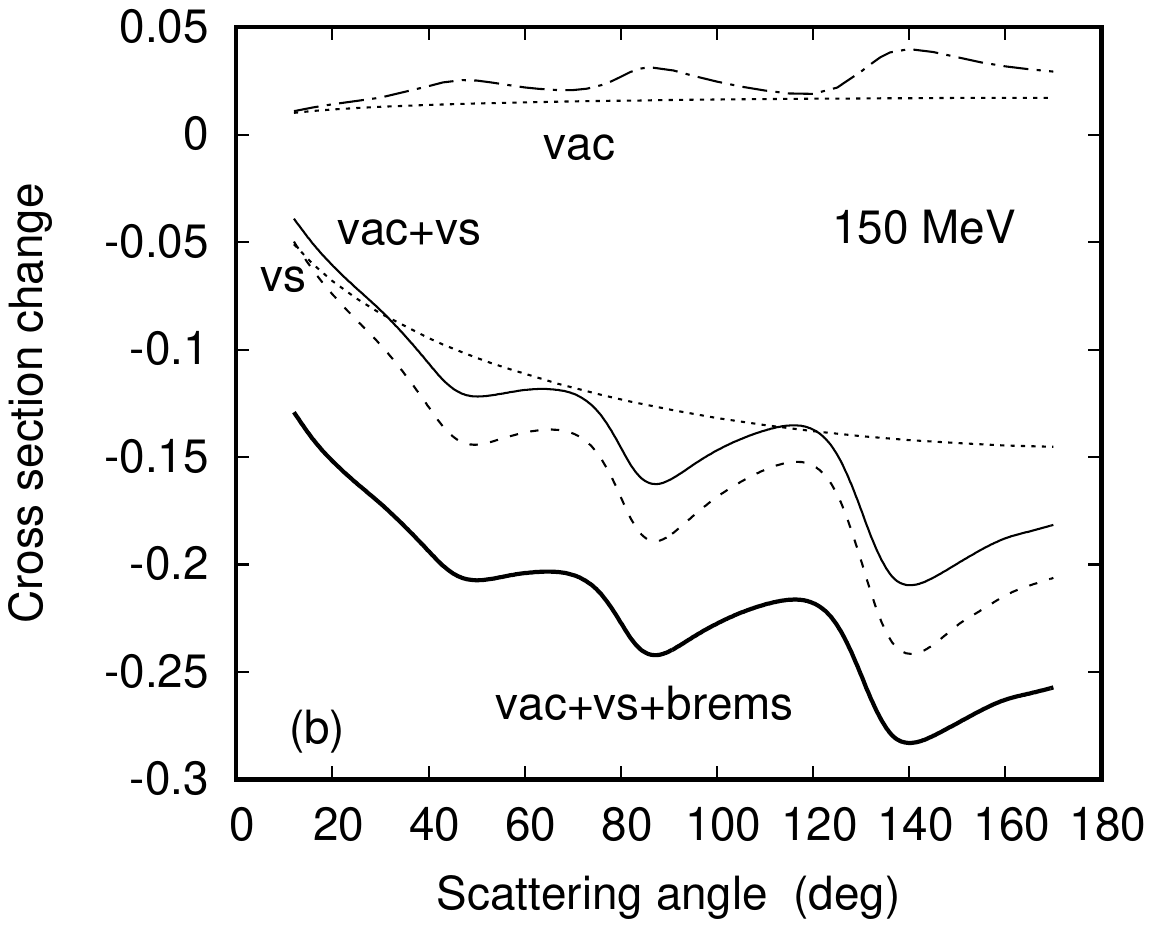}
\caption
{
Cross section change $\Delta \sigma^C$ in (a) 56 MeV and (b) 150 MeV $e + ^{208}$Pb collisions as a function of scattering angle $\vartheta_f$.
Shown are the results for vacuum polarization ($-\cdot - \cdot -)$, vertex and self-energy correction $(----)$ and for both\\ (-----------, thin line),
as well as the additional inclusion of the soft-bremsstrahlung contribution for $\omega_0=1$ MeV (-----------, thick line). Also shown are the Born results $\Delta \tilde{\sigma}^{\rm vac}$ ($\cdots\cdots$, upper line) and $\Delta \tilde{\sigma}^{\rm vs}$ ($\cdots\cdots$, lower line).
}
\end{figure}

\subsection{The $^{208}$Pb nucleus}

Fig.4 provides the angular distribution of the Coulombic cross section at the impact energies 56 and 150 MeV, as well as for 167 MeV where experimental data are available,
which were measured relative to $^{12}$C and normalized to the $^{12}$C phase-shift theory \cite{Fr72}.
For the extended lead nucleus, diffraction oscillations are already present at 150 MeV, while having  a still  earlier onset (i.e. at smaller angles) at 167 MeV.

The QED changes in the differential cross section are plotted in Fig.5, again in comparison with the Born results from (\ref{3.1}).
The deviations between the two prescriptions are considerably larger than for $^{12}$C, even at 56 MeV (Fig.5a), with notable differences already at the smallest angles.
Since the Born results $\Delta\tilde{\sigma}^{\rm vac}$ and $\Delta \tilde{\sigma}^{\rm vs}$ coincide with those from Fig.2, the effect of Coulomb distortion when proceeding from $^{12}$C to $^{208}$Pb becomes obvious.
We note that the combined inclusion of $U_e$ and $V_{\rm vs}$, leading to $\Delta \sigma^{\rm vac+vs}$, differs from the sum resulting from  the  separate treatments, $\Delta \sigma^{\rm vac} + \Delta \sigma^{\rm vs}$.
At 56 MeV, this difference is up to 3\% (in comparison to 1\% for $^{12}$C), increasing with energy.

\begin{figure}
\vspace{-1.5cm}
\includegraphics[width=11cm]{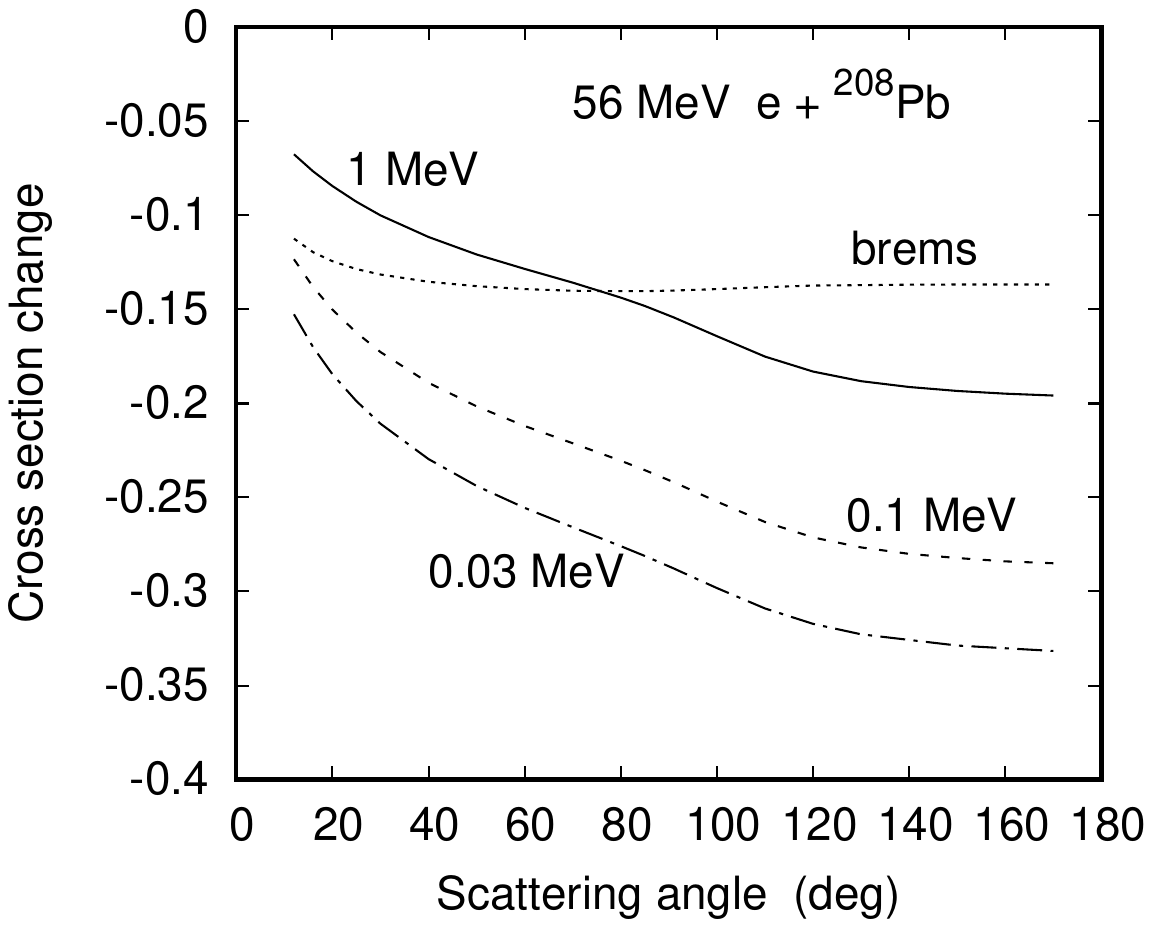}
\vspace{-0.5cm}
\caption
{
Angular dependence of the nonperturbative cross-section change by all QED effects for 56 MeV electrons colliding with $^{208}$Pb
and cut-off frequencies $\omega_0=1$ MeV (------------), 0.1 MeV $(----)$ and 0.03 MeV $(-\cdot -\cdot -$).
Also shown is the isolated contribution from soft bremsstrahlung for $\omega_0=0.1$ MeV $(\cdots\cdots)$.
}
\end{figure}

For 150 MeV (Fig.5b), the nonperturbative QED effects show oscillations, the minima of which correspond to the minima in the respective differential cross section.
In a similar way as for $^{12}$C, the cross-section modifications are particularly large and thus easily discernable when the Coulombic cross section does not notably change with angle (which is the case in the region of a diffraction minimum).

Fig.6 provides the influence of the soft bremsstrahlung when the detector resolution is changed.
The brems\-strahlung itself is approximately constant in angle at 56 MeV,
apart from the foremost regime. It increases, however, in strength when the cut-off frequency $\omega_0$ is lowered,
according to the logarithmic dependence $(\ln 2\omega_0/c^2$) in the formula (\ref{2.4}).

\begin{figure}
\vspace{-1.5cm}
\includegraphics[width=11cm]{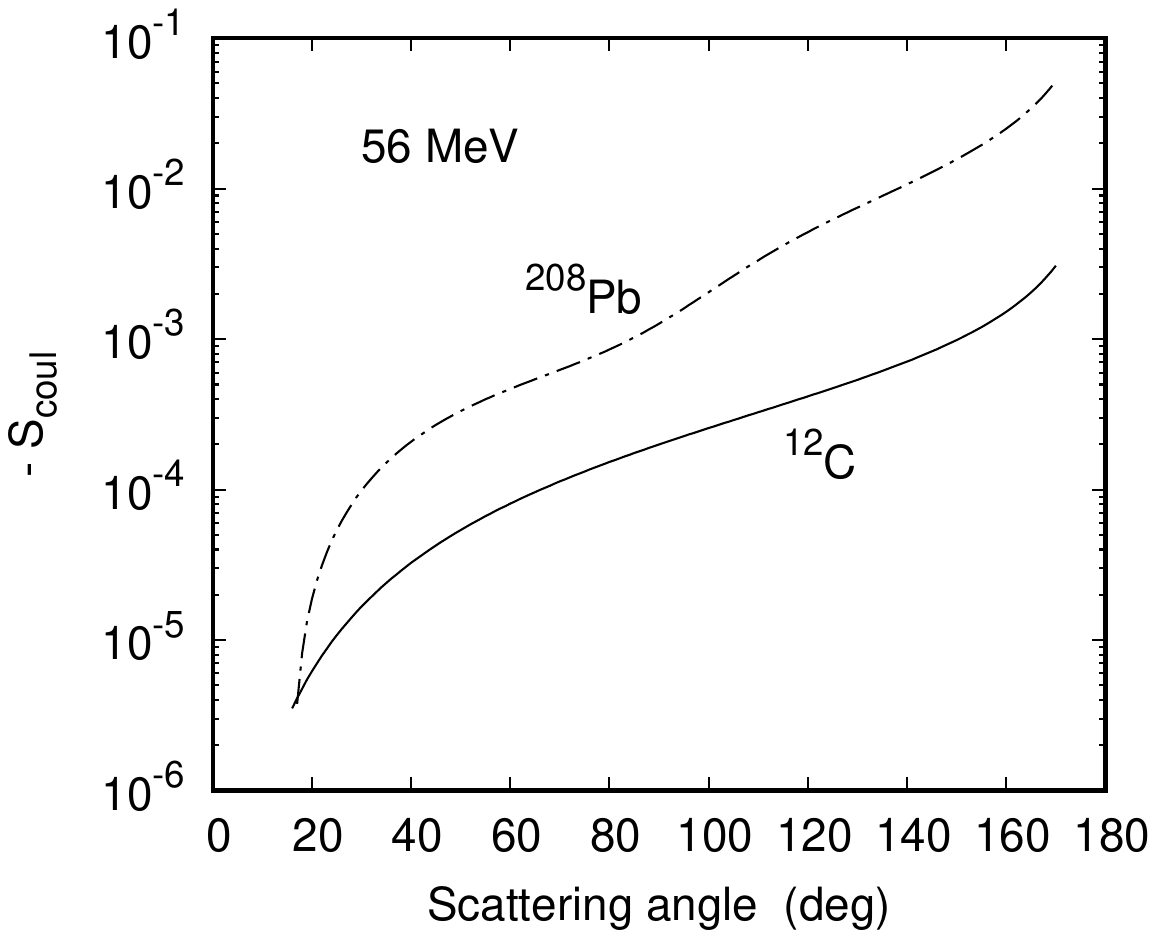}
\vspace{-0.5cm}
\caption
{
Coulombic spin asymmetry $- S_{\rm coul}$ for 56 MeV electrons colliding with $^{12}$C (----------) and $^{208}$Pb $(-\cdot - \cdot -$) as a function of scattering angle $\vartheta_f$.
}
\end{figure}

\subsection{Spin asymmetry}

For perpendicularly polarized incident electrons, the Sherman function $S$ is defined as the relative cross section difference when the initial spin is flipped from up ($\uparrow$) to down ($\downarrow$),
\begin{equation}\label{3.8}
S\;=\;\frac{d\sigma/d\Omega_f(\uparrow) - d\sigma/d\Omega_f(\downarrow)}{d\sigma/d\Omega_f(\uparrow)+d\sigma/d\Omega_f(\downarrow)}.
\end{equation}
Correspondingly, the modification of the spin asymmetry  by the radiative corrections, relative to the Coulombic Sherman function $S_{\rm coul}$, is calculated from
\begin{equation}\label{3.9}
d\,S\;=\;\frac{S}{S_{\rm coul}}\;-\,1.
\end{equation}
For higher collision energies when diffraction induces zeros in $S_{\rm coul}$, the definition (\ref{3.9}) is no longer meaningful. Therefore we have restricted the spin investigations to an energy of 56 MeV.
For this energy, the Coulombic spin asymmetry is displayed in Fig.7 for both targets.
A logarithmic scale is used (and hence $-S_{\rm coul}$ is shown) to demonstrate the strong increase of the spin asymmetry with scattering angle.
It should be noted that $S$ is much larger for $^{208}$Pb, since the spin asymmetry increases with nuclear charge due to the stronger relativistic effects in the 
electron-nucleus encounter.

\begin{figure}
\vspace{-1.5cm}
\includegraphics[width=11cm]{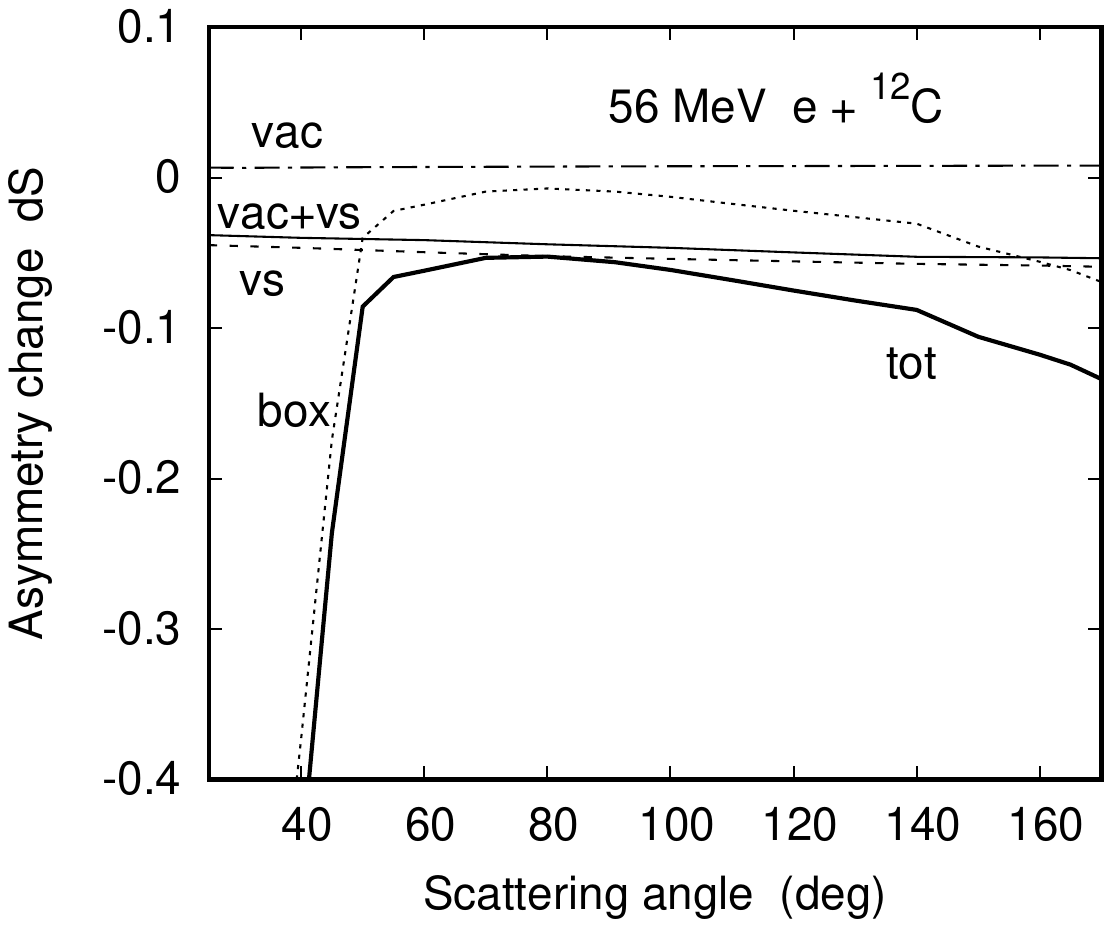}
\vspace{-0.5cm}
\caption
{
Change $dS$ of the Sherman function from 56 MeV $e+^{12}$C collisions by the nonperturbative QED effects and dispersion (------------, thick line) as a function of scattering angle $\vartheta_f$.
Also shown are the separate contributions from vacuum polarization ($dS_{\rm vac}, \;-\cdot -\cdot -$), from the vs correction ($dS_{\rm vs},\;----$), from both ($dS_{\rm vac+vs}$, ------------- thin line) and from dispersion ($dS_{\rm box},\;\cdots\cdots$).
}
\end{figure}

Fig.8 depicts the change $dS$ for $^{12}$C by means of the QED effects and the dispersion correction.
Like in the case of the cross-section modifications, the effect of vacuum polarization is small, at most 1\%. The vs contribution is of opposite sign
and  considerably larger in magnitude, well beyond the factor of $2$ anticipated from exact bound-state QED investigations \cite{Sh00}.
We recall that the spin-asymmetry change by the QED effects is zero in the Born approximation (\ref{2.1}) or (\ref{3.1}), since the leading-order cross section
is only multiplied by a factor which drops out in (\ref{3.8}). Moreover, as the soft-bremsstrahlung contribution contains the leading-order cross section as a factor, it does also not add to any asymmetry change 
in a higher-order approach. This was already stated by Johnson et al \cite{Jo62}, who calculated the QED corrections to $S$ within the second-order Born approximation in the Coulomb field.
In this context our previous Born results for $dS_{\rm vsb}$ \cite{Jaku21} should only be considered as qualitative estimates, since Coulomb distortion was not included in the contributions from vs and from soft bremsstrahlung.

Also shown in Fig.8 is the asymmetry change from dispersion \cite{Jaku22}, which tends to large negative values for small angles.
The Sherman function with inclusion of all radiative corrections can be estimated by
\begin{equation}\label{3.10}
S_{\rm tot}\;\approx\;S_{\rm vac+vs} \,+\,d S_{\rm box} \,S_{\rm coul}\,\frac{d\sigma_{\rm coul}/d\Omega_f}{d\sigma^{\rm QED}/d\Omega_f},
\end{equation}
where $S_{\rm vac+vs}$ and $S_{\rm coul}$ are calculated from the leading-order term in (\ref{3.3}) and (\ref{3.1}), respectively, or alternatively in terms of the direct (A) 
 and spin-flip (B) amplitudes as obtained from the phase-shift analysis,
\begin{equation}\label{3.11}
S\;=\;\frac{2 \mbox{ Re } \{A B^\ast\}}{|A|^2 +|B|^2}.
\end{equation}

\begin{figure}
\vspace{-1.5cm}
\includegraphics[width=11cm]{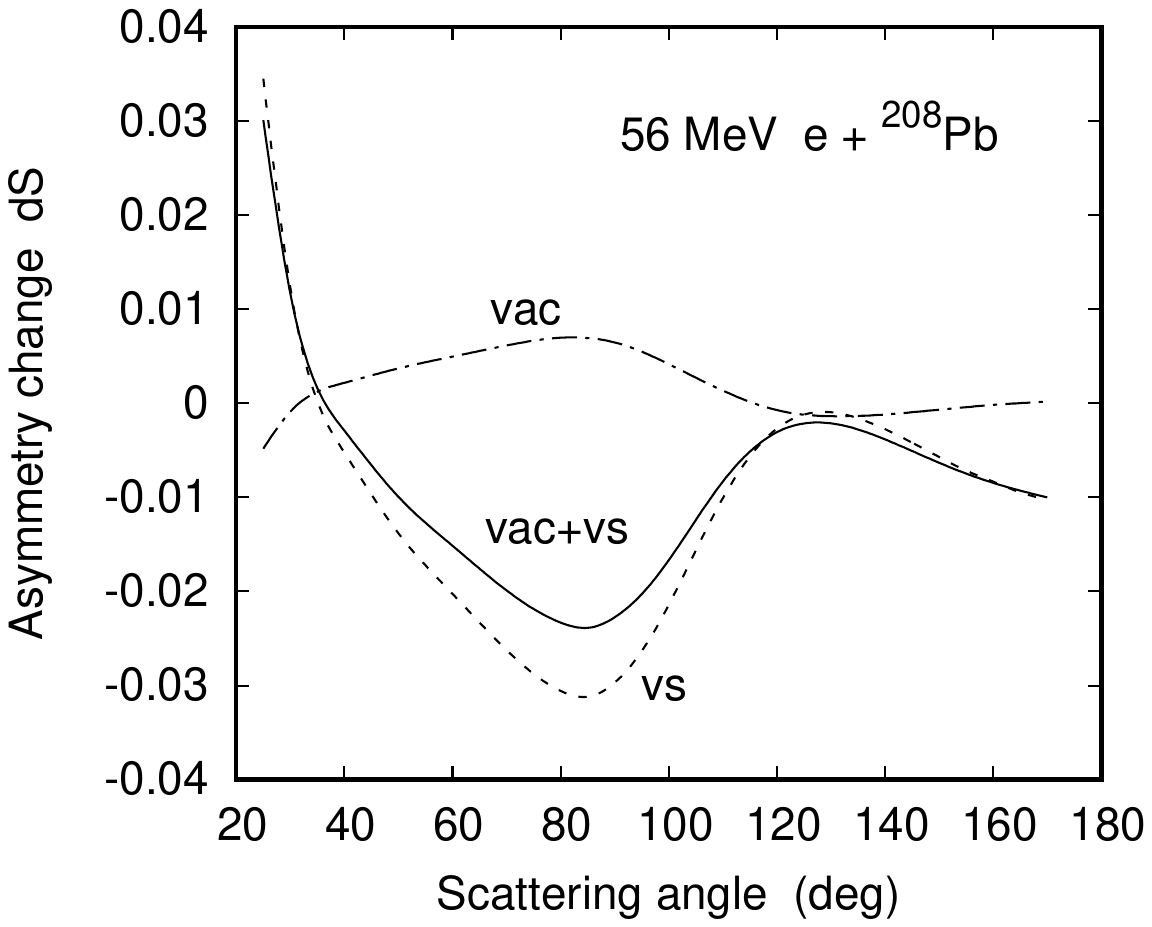}
\vspace{-0.5cm}
\caption
{
Change $dS$ of the Sherman function from 56 MeV $e+^{208}$Pb collisions by the nonperturbative QED effects ($dS_{\rm vac+vs}$, ------------) as a function of scattering angle $\vartheta_f$.
Included are the separate contributions from vacuum polarization ($dS_{\rm vac},\;-\cdot -\cdot -$) and from the vs effect ($dS_{\rm vs},\;----$).
}
\end{figure}

Furthermore, $d\sigma^{\rm QED}/d\Omega_f$ is calculated from (\ref{3.3}) by omission of dispersion, while $S_{\rm box}$ (and consequently $dS_{\rm box}$) is obtained from (\ref{3.1}) by dropping all three QED contributions.
Into the formula (\ref{3.10}) enters the assumption that the cross-section change due to dispersion is small (for a $^{12}$C target, it is below 1\% for collision energies up to 100 MeV),
such that it can be omitted in the denominator.
Consequently, the total change of asymmetry can be found from
$$dS_{\rm tot}\;=\;\frac{S_{\rm tot}}{S_{\rm coul}}\,-1$$
\begin{equation}\label{3.12}
\approx\; dS_{\rm vac+vs}\, + \,dS_{\rm box}\;\frac{1}{1+(\Delta \sigma^{\rm vac+vs}+\Delta \sigma^{\rm soft})},
\end{equation}
which is also displayed in the figure.
Actually for a low-$Z$ target like $^{12}$C the determination of the asymmetry changes suffers from large numerical instabilities, 
 which are partly smoothed in the figure.

In Fig.9 the spin-asymmetry change by the QED effects is displayed for a $^{208}$Pb target. Diffraction effects are already perceptible at 56 MeV, producing a zero in $S_{\rm coul}$ near $\vartheta_f=16^\circ$, which induces the strong rise of $dS$ near the smallest angles shown in the figure.
Also, as compared to the nearly constant values of $dS_{\rm vac}$ or $dS_{\rm vs}$ for a carbon target at the same energy,
strong angular variations of the QED effects take place for lead, although the Coulombic cross section shows hardly any modulations.
However, the total QED spin-asymmetry changes are smaller than the respective  changes for $^{12}$C.
It is noteworthy that even slight diffraction effects cause sign changes in $dS_{\rm vac}$ and $dS_{\rm vs}$, such that there is an angular region (at 56 MeV between $120^\circ$ and $160^\circ$) where both QED modifications are of the same sign.

\begin{figure}
\vspace{-1.5cm}
\includegraphics[width=11cm]{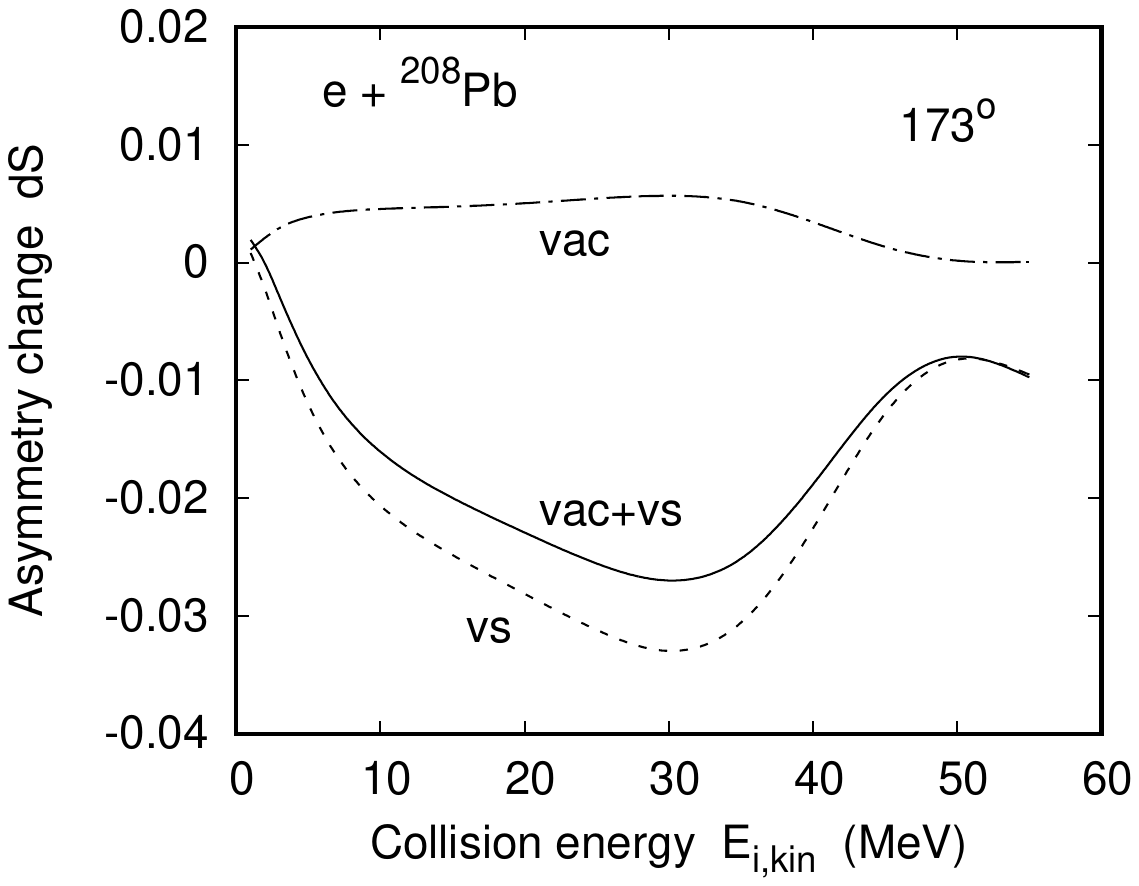}
\vspace{-1.0cm}
\caption
{
Change $dS$ of the Sherman function for $e+^{208}$Pb collisions at $\vartheta_f=173^\circ$ by the QED effects (---------------) as a function of collision energy $E_{\rm i,kin}=E_i-c^2$.
Included are the contributions $dS_{\rm vac}$ from vacuum polarization $(-\cdot - \cdot -$) and $dS_{\rm vs}$ from the vertex and self-energy correction $(----$).
}
\end{figure}

The behaviour of the Sherman function at very low collision energies is interesting from an experimental point of view
in the context of accuracy tests of different kinds of detectors.
 We provide in Fig.10 the energy dependence of the QED corrections at an angle of $173^\circ$, used in a recent precision experiment where 5 MeV electrons collided with a gold target \cite{Gr20}.
At this angle, $dS_{\rm vac}$ first increases with energy up to 0.5\%, and then decreases again beyond 30 MeV.
The vs contribution shows the opposite behaviour, with $|dS_{\rm vs}/dS_{\rm vac}|$
increasing from 1.5 at 3 MeV to 10 or more at the largest energies considered.

The similarity between the energy pattern (Fig.10) and the angular pattern (Fig.9) of the nonperturbative QED corrections indicates their basic dependence on the momentum transfer, $|\bfq| \approx 2 k_i \sin (\vartheta_f/2)$, into which the two variables enter as product.

\section{Conclusion}

The QED corrections to the elastic scattering cross section and to the beam-normal spin asymmetry were estimated by using a nonperturbative approach
in terms of  a suitable potential  for the vertex and self-energy correction, and the Uehling potential for vacuum polarization.
When investigating electron scattering from the $^{12}$C nucleus, notable deviations from the respective Born predictions for the
cross-section change were only found near and above 150 MeV impact energy, which are increasing with scattering angle.
In particular, the correction by the vs contribution, although mostly of opposite sign as compared to the effect of vacuum polarization,
is in magnitude considerably larger than the factor of two hitherto assumed from the results of exact low-energy bound-state considerations.

Like the cross-section changes by dispersion (estimated with or without the use of a closure approximation), the nonperturbative QED results show an oscillatory behaviour near the diffractive cross-section minima.
The numerical accuracy of our estimated QED cross-section changes is better than 0.5\% at 56 MeV, deteriorating to 5\% at 238 MeV for the backmost angles.

In case of the lead target, the deviations from the Born QED results are quite large, up to nearly a factor of 2 at backmost angles
even for a low energy of 56 MeV.
A diffraction pattern  emerges at energies near 100 MeV, with an increasing number of structures at higher energies, in concord with the diffractive structures of the Coulombic cross section. 
The numerical accuracy is higher than for $^{12}$C, below 1\% even at 150 MeV.
One has to keep in mind that the size of the total QED corrections depends strongly on the contribution of the soft bremsstrahlung, which in turn is controlled by the resolution of the electron detector.

The nonperturbative consideration of the vs effect allows also for a consistent estimate of the Sherman function. For low collision energies, its changes by the QED effects increase strongly with energy.
For lead this holds up to about 30 MeV at backward angles which are of particular interest to the experimentalists due to the large values of the spin asymmetry.
For example, at $170^\circ$ and 3.5 MeV, these QED  changes amount to  $dS\approx -0.5\%$, while at 5 MeV, $dS \approx -0.9\%$ for both targets.
On the other hand, at 56 MeV, they are about 5\% for $^{12}$C and somewhat less (at most 3\%) for $^{208}$Pb in the whole angular regime.
The numerical accuracy of $dS$ for carbon is unfortunately quite poor, partly due to the small absolute values of $S$ (in the forward regime), and partly due to numerical instabilities when solving  the Dirac equation
(in the backward hemisphere).
It amounts up to 0.5\% at 3 MeV and 3\% at 10 MeV, but $10-15\%$ at 56 MeV.
For lead, the results are stable, with an accuracy of $\lesssim 0.25\%$ at 30 MeV and $\lesssim 1\%$ at 56 MeV.

The dispersion effects on the cross section are small, but on the Sherman function they are formidable, even for 56 MeV electron impact on $^{12}$C. They lead to a total change of $S$ by the radiative corrections up to 50\% or more at the smallest angles.
An investigation of dispersion for a lead target is in progress.

\vspace{0.5cm}

\noindent{\large\bf Acknowledgments}

I would like to thank C.Sinclair for directing my interest to his work.


\vspace{1cm}

\end{document}